\documentstyle[%
aps,twocolumn,epsfig%
]{revtex}
\begin{document}
\draft

\twocolumn[\hsize\textwidth\columnwidth\hsize\csname
@twocolumnfalse\endcsname

\title{Importance of Correlation Effects on Magnetic Anisotropy in Fe and Ni }

\author{Imseok Yang, Sergej Y. Savrasov, and Gabriel Kotliar
}
\address{Department of Physics and Astronomy and Center for Condensed Matter Theory, 
Rutgers University, Piscataway, NJ 08854
}

\date{\today}
\maketitle

\begin{abstract}
We calculate magnetic anisotropy energy of Fe and Ni by taking into
account the effects of strong electronic correlations, spin-orbit coupling, 
and non-collinearity of intra-atomic magnetization. The LDA+U method is used
and its equivalence to dynamical mean--field theory
in the static limit is emphasized. Both experimental magnitude of MAE and direction
of magnetization are predicted correctly near $U=4\ eV$ for  Ni and
$U=3.5\ eV$ for  Fe.   Correlations modify  
one--electron spectra which are now in better agreement
with experiments. 
\end{abstract}


]

The calculation of the magneto-crystalline anisotropy energy 
(MAE)~\cite{vanVleck:37,brooks,fletcher,sloncewskij,asdente}  
of  magnetic materials containing transition-metal elements
from first principles electronic structure calculations is
a long-standing problem. The MAE is defined as the difference 
of total energies with the orientations of   
magnetization pointing in different, e.g., (001) and (111), crystalline axis. 
The difference is not zero because of spin-orbit effect, 
which couples the magnetization to the lattice, and 
determines the direction of magnetization, called the easy axis.
 
Being a ground state property, the MAE should be accessible in principle via 
density functional theory (DFT)~\cite{hk64,ks65}. 
Despite the primary difficulty related
to the smallness of MAE ($\sim 1\ \mu eV/$atom),
great efforts to compute the quantity
with advanced total energy methods based on local density
approximation (LDA) combined with the development of
faster computers, have seen success in predicting its correct orders of 
magnitudes~\cite{efn,halilov,jansen,tjew,dks}. 
However, the correct easy axis of  Ni has not been predicted by
this method and the fundamental problem of understanding MAE is still
open. 

A great amount of work has been done to understand what is the difficulty
in predicting the correct axis for Ni. Parameters within the LDA calculation 
have been varied to capture physical effects which might not be correctly
described. These include (i) scaling spin-orbit coupling in order 
to enlarge its effect on the MAE~\cite{halilov,jansen}, (ii) calculating 
torque to avoid comparing large numbers of energy~\cite{jansen},
(iii) studying the effects of the second Hunds rule in the
orbital polarization theory ~\cite{tjew}, (iv) analyzing
possible changes in the position of the Fermi level by 
changing the number of valence electrons ~\cite{dks},  (v)
using the state tracking method~\cite{freeman}, 
and (iv) real space approach~\cite{beiden}.

In this paper we take a new view that the
correlation effects within the $d$ shell are important for the magnetic anisotropy  
of  $3d$ transition metals like Ni. These effects are not captured by the LDA 
but are described by Hubbard--like interactions presented in these systems 
and need to be treated by  first principles  methods\cite {anisimov}.

Another effect which has
not been  investigated  in the context of  magnetic anisotropy calculations
is the non-collinear nature of intra-atomic magnetization~\cite{singh}.
It is expected to be important when spin-orbit coupling and correlation effects 
come into play together. In this article 
we show that when we include these new ingredients into the calculation we solve
the long-standing problem of predicting the correct easy axis of  Ni.

We believe that the physics of transition metal compounds
is intermediate between atomic limit where the localized
$d$ electrons are treated in the real space and fully itinerant limit 
when the electrons are described by band theory in k space.
A many--body method incorporating these two important limits 
is the dynamical mean--field theory (DMFT)~\cite{gabi:rmp}.
The DMFT approach 
has been extensively used to study model Hamiltonian of correlated electron systems
in the weak, strong and intermediate coupling regimes. It has
been very successful in describing the physics of realistic 
systems, like the transition metal oxides and, therefore, 
is expected to treat properly the materials with $d$ or $f$ electrons.

Electron-electron
correlation matrix $U_{\gamma _{1}\gamma _{2}\gamma _{3}\gamma
_{4}}=\left\langle m_{1}m_{3}\left| v_{C}\right| m_{2}m_{4}\right\rangle
\delta _{s_{1}s_{2}}\delta _{s_{3}s_{4}}$ for $d$ orbitals is the 
quantity which takes strong correlations into account. 
This matrix can be expressed via Slater
integrals $F^{(i)} $,  $ i=0,2,4,6$  in the standard manner.
The inclusion of this interaction generates self--energy 
$\Sigma_{\gamma_1\gamma_2}(i\omega_n,\vec{k})$ 
on top of the one--electron spectra.
Within DMFT it is approximated by momentum independent 
self--energy $\Sigma_{\gamma_1\gamma_2}(i\omega_n)$.

A central quantity of the dynamical mean--field theory is the 
one--electron on--site Green function
\begin{eqnarray}
G_{\gamma_1\gamma_2}(i\omega_n)=&\sum_{\vec{k}}&\left[(i\omega_n+\mu) 
O_{\gamma_1\gamma_2}(\vec{k}) 
-H^0_{\gamma_1\gamma_2}(\vec{k})\right. \nonumber \\
&+&\left. v_{dc}-\Sigma_{\gamma_1\gamma_2}
(i\omega_n)\right]^{-1}.  \label{dmft}
\end{eqnarray}
where $H^0_{\gamma_1\gamma_2}(\vec{k})$ is the one-electron Hamiltonian
standardly treatable within the LDA.
Since the latter already includes the electron-electron interactions
in some averaged way, we subtract the double counting term $v_{dc}$~\cite{laz}.
The use of realistic localized orbital representation such as 
linear muffin--tin orbitals \cite{OA75} leads us to 
include overlap matrix $O_{\gamma_1\gamma_2}(\vec{k})$ into the calculation. 

The DMFT reduces the problem to solving effective impurity
model where the correlated $d$ orbitals are treated as an impurity level hybridized
with the bath of conduction electrons. The role of hybridization is
played by the so--called bath Green function defined as follows:
\begin{equation}
[{\cal G}_0^{-1}]_{\gamma_1\gamma_2}(i\omega_n)=
G_{\gamma_1\gamma_2}{}^{-1}(i\omega_n)
+\Sigma_{\gamma_1\gamma_2}(i\omega_n).
\end{equation}
Solving this impurity model gives access to 
the self--energy $\Sigma_{\gamma_1\gamma_2}(i\omega_n)$ for
the correlated electrons. The one--electron Green function~(\ref{dmft}) is now
modified with new $\Sigma_{\gamma_1\gamma_2}(i\omega_n)$, 
which generates a new bath Green function. Therefore,
the whole problem requires self--consistency.

In this paper we confine ourselves to zero temperature and make
an additional assumption on solving the impurity model
using the Hartree--Fock approximation.  
In this approximation the self--energy reduces to
\begin{equation}
\Sigma_{\gamma_1\gamma_2}=\sum_{\gamma_3\gamma_4}
(U_{\gamma _{1}\gamma _{2}\gamma
_{3}\gamma _{4}}-U_{\gamma _{1}\gamma _{2}\gamma _{4}\gamma _{3}})
\bar{n}_{\gamma _{3}\gamma _{4}} \label{HF}
\end{equation}
where $\bar n_{\gamma_1\gamma_2}$ is the average occupation matrix for
the correlated orbitals. The off-diagonal elements of the occupancy matrix 
are not zero when spin-orbit coupling is included~\cite{sol}. 
The latter can be implemented following the prescription  of
Andersen~\cite{OA75} or more recent one by Pederson~\cite{pk}.

In the Hartree--Fock limit the self--energy is frequency independent and real. 
The self--consistency condition of DMFT can be expressed in terms of the 
average occupation matrix: Having started from 
some $\bar n_{\gamma_1\gamma_2}$ 
we find $\Sigma_{\gamma_1\gamma_2}$ according to~(\ref{HF}). Fortunately, the computation
of the on--site Green function~(\ref{dmft}) needs {\em not} to be performed. 
Since the self--energy is real, the new occupancies can be calculated from
the eigenvectors of the one--electron Hamiltonians with $\Sigma_{\gamma_1\gamma_2}-v_{dc}$
added to its $dd$ block. The latter can be viewed as an orbital--dependent
potential which has been introduced by the LDA+U method \cite{anisimov}.

The LDA$+$U method 
has been very successful compared with experiments at zero temperature in
ordered compounds. By establishing its equivalence to the static limit of the DMFT
we see clearly that dynamical mean--field
theory is a way of improving upon it, which is crucial for finite
temperature properties.

In this work we study the effect of the Slater parameter $F_0$
that is the Hubbard on site interaction U on the
magnetic anisotropy energy. The latter is calculated by taking the difference
of two total energies with different directions of
magnetization (MAE=$E(111)-E(001)$). The total energies are obtained via
fully self consistent solutions. Since the total energy
calculation requires high precision,  full potential
LMTO method~\cite{Sav} has been employed. For the $\vec{k}$ space
integration, we follow the analysis given by Trygg and co--workers~\cite{tjew} 
and use the special point method~\cite{froyen} with a
Gaussian broadening~\cite{mp} of $15\ mRy$.  
The validity of this procedure has been tested in their work ~\cite{tjew}.
For convergence of the total energies within desired accuracy, 
about $15000\ k$-points are needed. We used $28000\
k$-points to reduce possible numerical noise. 
Our calculations include non-spherical terms of the
charge density and potential both within the atomic spheres and in
the interstitial region~\cite{Sav}. All low-lying semi-core states
are treated together with the valence states in a common
Hamiltonian matrix in order to avoid unnecessary uncertainties.
These calculations are spin polarized and assume the existence of
long-range magnetic order. Spin-orbit coupling is implemented 
according to the suggestions by Andersen~\cite{OA75}.
We also treat magnetization as a general vector field, which realizes non-collinear
intra-atomic nature of this quantity.
Such general magnetization scheme has been recently discussed in ~\cite{singh}.

We now discuss our calculated MAE. We first test our method in case
of  LDA ($U=0$). To compare with previous calculations, we  
turn off the non-collinearity of magnetization which makes it 
collinear with the quantization axis.
The calculation gives correct orders of magnitude for
both  fcc Ni and  bcc Fe but with the
wrong easy axis for  Ni, which is the
same result as the previous result~\cite{tjew}. 
Turning on the non-collinearity results  
in a a larger value of the absolute value of the MAE
($2.9\ \mu eV$) for  Ni but the easy axis predicted to be (001) which is still wrong. The magnitude of the  
experimental MAE of Ni is  $2.8\ \mu eV$ aligned along
$(111)$ direction~\cite{landolt}.

\begin{center}
\begin{figure}
        \epsfig{file=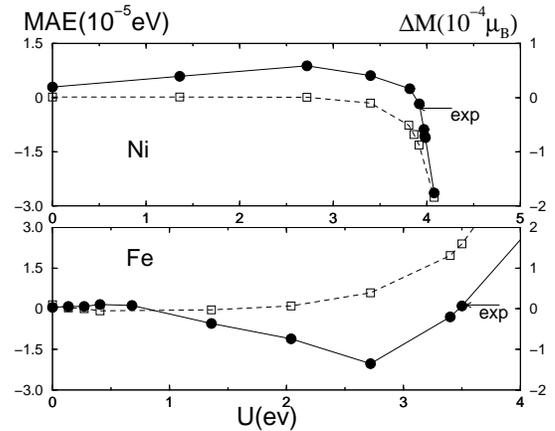, width=0.4\textwidth}
        \vspace{6pt}
        \caption{Experimental and calculated magnetocrystalline
anisotropy energy $\mbox{MAE}=E(111)-E(001)$ (filled circle)
and the difference of magnetic
moment $\Delta M=M(001)-M(111)$ (square) for  Fe (top) and 
        Ni(bottom). Experimental
 MAEs are marked by arrows for  Fe ($1.4\ \mu eV$) and  Ni ($-2.8\ \mu eV$).
 }
\label{mae}
\end{figure}
\end{center}

We now describe the effect of strong correlations,
which is crucial in predicting the correct axis of Ni (see Fig.\ \ref{mae}). 
As U increases, the MAE of Ni
smoothly increases until $U$ reaches $2.5\ eV$ and then smoothly decreases
up to the value $3.8\ \mu eV$. Around $U=3.9\ eV$, the MAE decreases abruptly to
negative value. Around $U=4.0\ eV$, the experimental order of
magnitude and the correct easy axis (111) are restored.
The change from the wrong easy axis to the
correct easy axis occurs over the range of $\delta U\sim 0.2 eV$,
which is the order of spin-orbit coupling constant ($\sim 0.1
eV$).

For  Fe, the MAE decreases on increasing $U$ to negative values, where the magnetization
takes the wrong axis. From $U=2.7\ eV$, it increases back to the
correct direction of easy axis (positive MAE). Around $U=3.5\ eV$, it restores
the correct easy axis and the experimental value of MAE is reproduced.

It is remarkable that the values of $U$ necessary to reproduce
the correct magnetic anisotropy energy are very close
to the values which are needed to  describe photoemission 
spectra of these materials~\cite{kl99}. This
shows an internal consistency of our approach and emphasizes the importance
of correlations.
 
We find direct correlation between the dependency of the MAE as a function of $U$ and 
the difference of magnetic moments ($\Delta M=-(M(111)-M(001)$) behaving
similarly. For  Ni, the difference of magnetic moments
is nearly $U$ independent up to $U=3\ eV$. For large $U$, 
it smoothly decreases from the positive value to the negative one.
It also decreases rapidly around $U=3.9\ eV$ in accord with MAE. For  Fe, the
difference is positive at $U=0$. It decreases
slightly to the negative values and then increases to the positive 
value over the range of $U < 2.7\ eV$ where MAE decreases.
For larger $U$'s, where MAE is coming back to positive value,
its slope is significantly larger than that at smaller $U$'s.

This concurrent change of MAE and the difference of magnetic moments 
suggests why some previous attempts based on force theorem~\cite{dks}
failed in predicting the correct easy axes. Force theorem
replaces the difference of the total energies by the difference of
one--electron energies. In this approach, the contribution from the
slight difference in magnetic moments does not
appear and, therefore, is not counted in properly. 
Unfortunately, we could not find any experimental data of 
magnetic moments to the desired precision ($10^{-4}\mu_B$) to compare with.  
We also have problems in reaching the convergence of the total energy
with desired accuracy for large values of $U$ in both Fe and Ni.

\begin{center}
\begin{figure}
        \epsfig{file=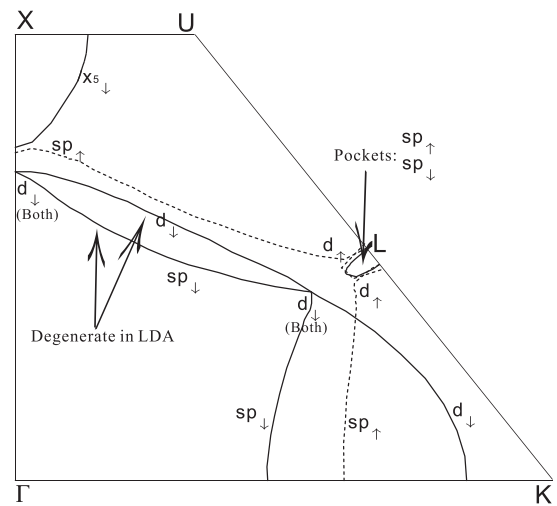, width=0.4\textwidth}
        \vspace{6pt}
        \caption{Calculated Fermi Surface of  Ni with the correlation effects
        taken into account.
        The solid and dotted lines
        correspond to majority and minority dominant spin carriers.
        Dominant orbital characters are expressed.
        Both experimentally confirmed $X_5$ pocket and $L$ neck can be seen.
        The $X_2$ pocket is missing, which is in agreement with
        experiments. There are two small $L$ pockets, which has
        not been found by experiments.}
\label{fermi}
\end{figure}
\end{center}

We now present implications of our results on the calculated electronic structure
for the case of Ni. One important feature which emerges from the calculation is the 
absence of the $X_2$ pocket (see Fig.\ \ref{fermi}). This has been
predicted by LDA but has not been found experimentally~\cite{wc}. The band
corresponding to the pocket is  pushed down well below the Fermi level. 
This is expected since correlation effects
are more important for slower electrons and the velocity
near the  pocket is rather  small.
It turns out that the whole band is submerged under the Fermi level.

There has been some suspicions that the incorrect position of  the
$X_2$ band within LDA was responsible for the incorrect prediction of the easy axis
within this theory. Daalderop and coworkers~\cite{dks} removed 
the $X_2$ pocket by increasing the number of valence electrons and 
found the correct easy axis. We therefore conclude that the absence of
the pocket is one of the central elements in determining the magnetic anisotropy,
and there is no need for any ad-hoc adjustment
within a theory which takes into account the correlations.

We now describe the effects originated from (near) degenerate states close
to the Fermi surface. These have been 
of primary interest in past analytic studies~\cite{Kondorskii,mori:74}.  
We will call such states {\em degenerate Fermi surface crossing} (DFSC) states.
The contribution to MAE by non-DFSC states comes from the fourth
order perturbation. Hence it is of the order of $\lambda^4$. 
The energy splitting between DFSC states due to spin-orbit coupling is of the
order of $\lambda$ because the contribution comes from the first
order perturbation. Using linear approximation of the dispersion
relation $\epsilon(\vec{k}\lambda)$, the relevant volume
in $k$-space was found of the order $\lambda^3$. Thus, these DFSC
states make contribution of the order of $\lambda^4$.
However, there may be accidentally DFSC states
appearing along a line on the Fermi surface, rather than at a point.
We have found this case in our LDA calculation for Ni.
Therefore the contribution of DFSC states is as important as 
the bulk non-DFSC states though the degeneracies occur only 
in small portion of the Brillouin zone.


The importance of the DFSC states leads us to comparative analysis
of the LDA and LDA+U band structures near the Fermi level. 
In LDA, five bands are crossing the Fermi level at nearly the same points
along the $\Gamma X$ direction.
Two of the five bands are degenerate for the residual symmetry and
the other three bands accidentally cross the Fermi surface at nearly the
same points. There are two $sp$ bands with spin up and spin down,
respectively. The other three bands are dominated by $d$ orbitals.
In LDA$+$U, one of the $d$ bands is pushed down below the Fermi surface.
The other four bands are divided into two degenerate pieces at the Fermi
level (see Fig. \ref{fermi}): 
Two symmetry related degenerate $d_\downarrow$ bands and
two near degenerate $sp_\uparrow$ and $sp_\downarrow$ bands. 
In LDA, we found that two bands are
accidentally near degenerate along the line on the Fermi surface
within the plane $\Gamma X L$. One band is
dominated by $d_\downarrow$ orbitals. The other is dominated by
$s_\downarrow$ orbitals near $X$ and by $d_\downarrow$ orbitals off $X$. 
In LDA+U, these accidental DFSC states disappear(see Fig. \ref{fermi}). 
Instead,
there is new two-fold DFSC states along $\Gamma
L$ direction, both of which are dominated by  $d_\downarrow$ orbitals.

As we have seen, the on-site repulsion
$U$ reduces the number of DFSC states along $\Gamma X$
direction while increasing that of DFSC states along $\Gamma
L$ direction.    
Based on the tight--binding model, Mori and
coworkers~\cite{mori:74} have shown that DFSC states along $\Gamma L$
direction result in the magnetization aligned  
along $(1,1,1)$ direction and 
DFSC states along $\Gamma X$ direction result in 
the magnetization aligned 
along $(0,0,1)$ direction. 
Since the strong correlation does precisely this, 
we conclude that disappearance of DFSC states
along $\Gamma X$ direction and their appearance along
$\Gamma L$ direction is another important element 
that determines the easy axis of Ni.

Unlike LDA, we have found two extra very tiny $L$ pockets in LDA$+$U
(see Fig. \ref{fermi}).
Both of them are dominated by $sp$ orbital with opposite spins. Being
small, these extra pockets may be artifacts of LDA$+$U.



To conclude, we have demonstrated that it is possible to perform
highly precise calculation of the total energy in order to obtain both
the correct easy axes and the magnitudes of MAE for Fe and Ni. 
This has been accomplished
by including the strong correlation effect via the Hubbard-like on-site repulsion
$U$ and incorporating the non--collinear magnetization. In
both Fe and Ni, the on-site $U$ takes physically acceptable
values consistent with the values known from atomic physics.
The calculations performed are state of the art in what can
currently be achieved for realistic treatments of correlated solids.
Further studies should be devoted to improving the quality of the
solution of the impurity model within DMFT and extending the
calculation to finite temperatures. 

This research was supported by the ONR grant No.\@ 4-2650.
GK would like to thank  K.  Hathaway for discussing the origin of  
magnetic anisotropy  and  G. Lonzarich for discussing dHvA data.
We thank R. Chitra for stimulating discussion.
We thank V. Oudovenko for setting up the computer cluster 
used to perform these calculations.
We have also used the supercomputer at the Center for Advanced 
Information Processing, Rutgers. IY thanks K. H. Ahn for discussions.


\end{document}